\documentstyle[preprint,aps]{revtex}
\begin{document}
\draft
\preprint{IFT-P.000/00}
\title{Fermion helicity flip in weak gravitational fields}
\author{ R.\ Aldrovandi, G.\ E.\ A.\ Matsas, S.\ F.\ Novaes,
and D.\ Spehler\cite{byline}}
\address{Instituto de F\'{\i}sica Te\'orica,
Universidade  Estadual Paulista, \\
Rua Pamplona 145, CEP 01405-900 S\~ao Paulo, Brazil.}
%
\maketitle
\begin{abstract}
The helicity flip of a spin-${\textstyle \frac{1}{2}}$ Dirac particle
interacting gravitationally with a scalar field is analyzed in the
context of linearized quantum  gravity.  It is shown that massive
fermions may have their helicity flipped by gravity, in opposition to
massless fermions which preserve their helicity.
\end{abstract}


The behavior of a spinning particle in a gravitational field is a
long standing problem \cite{P,W}, which has known recently a
revival of interest. In particular, whether gravity flips or not
the helicity of a spin-${\textstyle \frac{1}{2}}$ Dirac particle
has been examined lately in the semiclassical context
\cite{CP,A}. Although a final answer will have to
wait for a complete theory of quantum gravity, we believe the
linearized quantum approach can shed  new light on the subject.
Our aim in this letter is to examine the problem of fermion
helicity flip in this context. Namely, we calculate at tree level
the helicity flip probability for a spin-${\textstyle
\frac{1}{2}}$  Dirac particle interacting via a graviton exchange
with a massive scalar particle  in Minkowski space. We conclude
that only massive fermions can have their helicity flipped.  The
main virtue of our treatment is the fact that all concepts
employed are simple, and widely used in field theory.

The coupling of  matter fields with  gravity in the context of the
linearized theory is  obtained using  weak field approximation. In
this vein,  the spacetime metric may be written  as $g_{\mu\nu} =
\eta_{\mu\nu} + \kappa h_{\mu\nu}$,  where  $\eta_{\mu \nu}$ is the
Minkowski metric,  $\kappa \equiv \sqrt{32\pi G} = 8.211 \times
10^{-19}$ GeV$^{-1}$ plays the role of a small coupling constant, and
the perturbation  $h_{\mu \nu}$ represents the graviton field which is
supposed to be  quantized in Minkowski space using Cartesian
coordinates.  Hence, it is the metric tensor $\eta_{\mu \nu}$ which
will provide the canonical isomorphism between the vector space and
the corresponding dual space.

Next, by {\em minimally} coupling gravity to the scalar field $\phi$
\cite{DW},  and to the fermionic field $\psi$ \cite{ANS}, we obtain
the following interaction actions to first order in $\kappa$
\begin{equation}
S_{\phi \phi h} = \int \mbox{d}^4x \;\;
\frac{\kappa}{4} \left[h\left(\partial^{\alpha}\phi
\partial_{\alpha}\phi  -M^2  \phi^2 \right)
- 2 h^{\mu\nu} \partial_{\mu}\phi \partial_{\nu}\phi \right] \; ,
\label{sc:sc:h}
\end{equation}
and
\begin{eqnarray}
S_{\bar{\psi} \psi h}  = \int \mbox{d}^4x &&
\frac{\kappa}{2} \biggl\{ \left[  -\frac{i}{4} \bar{\psi} \left( \gamma^\nu
\partial_\mu + \gamma_\mu \partial^\nu \right) \psi + \frac{i}{4}
\left(\partial_\mu \bar{\psi} \gamma^\nu + \partial^\nu \bar{\psi} \gamma_\mu
\right) \psi \right] h_\nu^{\;\; \mu}
\nonumber \\
&+& \left( \frac{i}{2} \bar{\psi} \gamma^\nu \partial_\nu \psi
- \frac{i}{2} \partial_\nu \bar{\psi} \gamma^\nu \psi - m \bar{\psi} \psi
\right) h \biggr\} \; ,
\label{sp:sp:h}
\end{eqnarray}
where $M$, and $m$ are the scalar and fermion  masses
respectively, and $h \equiv h_\mu^\mu$.
The vertices associated with the couplings
$\phi(k_1)-\phi(k_2)-h^{\mu\nu}$ and
$\bar{\psi}(k_1)-\psi(k_2)-h^{\mu\nu}$ derived from the actions
(\ref{sc:sc:h}) and (\ref{sp:sp:h}) are
\begin{equation}
V_{\phi\phi h}^{\mu\nu} (k_1,k_2)=
i\frac{\kappa}{2} \left[ k_{1}^\mu  k_{2}^\nu
+ k_{1}^\nu k_{2}^\mu- \eta^{\mu\nu}
(k_1\cdot k_2 + m^2) \right] \; ,
\label{V:Fi:Fi:h}
\end{equation}
and
\begin{equation}
V_{\bar{\psi}\psi h}^{\mu\nu}  (k_1,k_2) =
i \frac{\kappa}{4} \left\{ \frac{1}{2}
\left[\gamma^\mu \left(k_1^\nu + k_2^\nu \right) +
\gamma^\nu \left(k_1^\mu + k_2^\mu \right)
\right] + \eta^{\mu\nu} \left[ (\not k_1 + \not k_2) - 2 m \right]
\right\} \; ,
\label{V:Psi:Psi:h}
\end{equation}
respectively.  In Eq. (\ref{V:Fi:Fi:h}) all the momenta are
considered incoming, whereas in Eq. (\ref{V:Psi:Psi:h}) they follow the
fermionic arrow.

We are now able to evaluate the helicity flip rate of a fermion ($F$)
interacting with a scalar ($S$) via a graviton exchange, {\it i.e.,}
$F (p) + S(q) \rightarrow F (k) + S(l)$ [see Fig. (\ref{fig:1})]. In
particular, the squared invariant amplitude for a initial left-handed
fermion, $\vert L \rangle$,  suffering a transition to a right-handed
one,  $\vert R \rangle$, due to a graviton exchange with a scalar
field is
\begin{equation}
N_{\vert L \rangle \to \vert R \rangle} =
\left| \bar{u}(k, {\cal S}_R) \; V_{\bar{\psi}\psi h}^{\mu\nu} (p,k)
\; u(p,{\cal S}_L) \;  D_{\mu\nu ,\alpha\beta}(p-k) \;
V_{\phi\phi h}^{\alpha\beta} (q,-l) \right|^2 ,
\label{N}
\end{equation}
where $D_{\mu\nu ,\alpha\beta}$ is the usual graviton propagator
\begin{equation}
D_{\mu\nu ,\alpha\beta} (p-k) = \frac{i}{2 (p-k)^2} \left( \eta_{\mu\alpha}
\eta_{\nu\beta} +  \eta_{\mu\beta} \eta_{\nu\alpha}
- \eta_{\mu\nu} \eta_{\alpha\beta} \right) \; .
\label{D}
\end{equation}
The fermion spinors satisfy
\[
u\left(p, {\cal S}_{L}\right) =
\frac{\left( 1 + \gamma_5 \gamma_\mu {\cal S}_{L}^\mu  \right)}{2}
u\left(p, {\cal S}_{L}\right) \; ,
\]
and
\[
u\left(k, {\cal S}_{R}\right) =
\frac{\left( 1 + \gamma_5 \gamma_\mu {\cal S}_{R}^\mu  \right)}{2}
u\left(k, {\cal S}_{R}\right)
\]
as usually,  where  we have introduced the polarization  four-vectors
\begin{eqnarray}
{\cal S}_{L}^\mu &=& -\frac{p^\mu}{m\beta_i}
+ \frac{\sqrt{1 - \beta_i^2}}{\beta_i} \; \eta^{\mu 0} \; ,
\nonumber \\
{\cal S}_{R }^\mu  &=& \frac{k^\mu}{m\beta_f}
- \frac{\sqrt{1 - \beta_f^2}}{\beta_f} \; \eta^{\mu 0}
\label{srl}
\end{eqnarray}
with $\beta_{i(f)}$ being the initial (final) fermion velocities.

Thus, letting (\ref{V:Fi:Fi:h}), (\ref{V:Psi:Psi:h}), and (\ref{D}) in
(\ref{N}) we obtain
\begin{eqnarray}
&&N_{\vert L \rangle \to \vert R \rangle} = \frac{\kappa^4}{512 t^2}
\nonumber \\
&\times& \biggl\{ \left[ 1 - \left({\cal S}_L \cdot {\cal S}_R \right) \right]
\left\{  (s - u)^2 \left[ (s - u)^2  - t (t - 4 M^2) \right]
- 16 m^2 M^2 \left[ M^2 (t - 4 m^2) +  (s - u)^2 \right] \right\}
\nonumber \\
&-& 4 \left[ \left( A \cdot {\cal S}_R \right) \left( k \cdot {\cal S}_L\right)
+ \left( A \cdot {\cal S}_L \right) \left( p \cdot {\cal S}_R \right)  \right]
\left[ (s-u)^2  + 2 t (t+3M^2) - 8 m^2 M^2  \right]
\nonumber \\
&+& 2 \left( p \cdot {\cal S}_R \right)   \left( k \cdot {\cal S}_L \right)
\left[4 (t+3M^2)^2(t-4m^2) +  (s-u)^2 (5t + 8 M^2) + 16 m^2
(t + 2M^2)^2\right]
\nonumber \\
&+& 8 \left( A \cdot {\cal S}_L \right) \left( A \cdot {\cal S}_R \right)  t
\biggr\} ,
\label{nr}
\end{eqnarray}
where  $s = (p+q)^2$, $t=(p-k)^2$, and $u=(p-l )^2$ are  the
Mandelstam variables, and  we have introduced the four-vector
\begin{eqnarray}
A^\mu &\equiv& (t + 3M^2) (k^\mu + p^\mu) -
\frac{1}{2} (s - u) (q^\mu  + l^\mu) \; .
\label{a}
\end{eqnarray}
In a similar way we can evaluate the squared  invariant amplitude for
a initially left-handed fermion to  remain with the same helicity
after the interaction.

The polarization of the scattered fermion can be measured by
\begin{equation}
P = 1 - \frac{ 2 N_{\vert L \rangle \to \vert R \rangle}}
	 {N_{\vert L \rangle \to \vert L \rangle}
	+ N_{\vert L \rangle \to \vert R \rangle}},
\label{P}
\end{equation}
where
\begin{eqnarray}
&&N_{\vert L \rangle \to \vert L \rangle} +
N_{\vert L \rangle \to \vert R \rangle}
= \frac{\kappa^4}{256 t^2}
\nonumber \\
&\times& \left\{  (s - u)^2 \left[ (s - u)^2  - t (t - 4 M^2) \right]
- 16 m^2 M^2 \left[ M^2 (t - 4 m^2) +  (s - u)^2 \right]
\right\}
\label{nlpnr}
\end{eqnarray}
depends only on the Mandestam variables and  the relevant masses.
Notice that $P=1$ indicates no depolarization (flip) of the initial
fermions, whereas $P=-1$ represents that all the left-handed initial
fermions were flipped.

In order to evaluate $P$, we can choose for instance the laboratory
frame where the scalar particle is initially at rest [$q^\mu = (M,
\vec{0})$]. In this frame, $s_{\text{lab}} = m^2 + M^2 + 2ME$,  and
$u_{\text{lab}} = 2 (m^2 + M^2) - s_{\text{lab}} - t_{\text{lab}}$
with
\[
t_{\text{lab}} = - 2M(E^2 - m^2)\left[ \frac{ M +  E \sin^2\theta
- \sqrt{M^2 - m^2 \sin^2 \theta} \; \cos \theta}{(E + M)^2 -
(E^2 - m^2)\cos^2\theta} \right] \; .
\]
Here $E$ is the initial fermion energy, and $\theta$  the fermion
scattering angle in this frame. The final expression for $P$ is rather
lengthy, and it is omitted here. In Fig. \ref{fig:2} we show the
behavior of $P$  as a function of the fermion mass. We have fixed the
initial fermion energy and the scalar particle mass, and shown the
curves for different values of the scattering angles.  Notice that the
larger the scattering angle, the larger the helicity flip. It is clear
from this figure that $P \to 1$ when $m \to 0$, {\it i.e.} there is no
helicity flip for massless fermions.  In Fig. \ref{fig:3} we present
$P$ as function of the scalar particle mass  at a fixed scattering
angle ($\theta= 30^\circ$), and for various values of the fermion
mass.

A simple expression for $P$ can be obtained by taking the limit of very
large scalar masses, ({\it i.e.,\/} $M \gg E > m$), which corresponds
to small squared transferred momentum approximation ($t \sim 0$). In
this regime,  we have ${\cal S}_{R}^\mu = - {\cal S}_{L}^\mu
\rightarrow p^\mu /m$, which leads to
\begin{equation}
 P \simeq  1 - \left( \frac{m^2}{E^2} \right) \; ,
\label{KEY}
\end{equation}
and makes clear that the helicity flip probability is proportional to
the square of the fermion mass. In this approximation, taking  for
instance a fermion with mass $\simeq 10$ eV, which is about the upper
bound for the  electron neutrino mass, with an energy in the range
$0.1$ -- $10$ MeV, we  obtain a helicity flip probablility of ${\cal
O}[10^{-(8 - 12)}]$.

Summarizing, we have shown that, in the context of linearized
quantum gravity, massive spin-${\textstyle \frac{1}{2}}$  Dirac
particles  have their helicity flipped due to gravitational
interaction. Notwithstanding, no helicity flip is expected for
massless  fermions. Our approach differs from the previous
semiclassical ones \cite{CP,A} in various aspects. The
gravitational interaction is described here by a massless spin-2
quantum field $h_{\mu \nu}$, and not by the spacetime curvature
as in the semiclassical approach.  Once assumed the reasonable
actions (\ref{sc:sc:h}) and (\ref{sp:sp:h}) obtained by minimally
coupling the matter fields to gravity, our results follow
straightforwardly. The fermion helicity flip appears as a
dynamical effect, coming from the  local coupling of spin to
gravity.

\acknowledgments

Three of us acknowledge  Conselho Nacional de Desenvolvimento
Cient\'{\i}fico e Tecnol\'ogico (CNPq) for partial (RA, SFN), and full
(GEAM) financial support, while DS is grateful to Funda\c{c}\~ao de
Amparo \`a Pesquisa do Estado de S\~ao Paulo (FAPESP) for full
financial support.

\begin{figure}
\protect
\caption{Feynman diagrams for the process $F(p) + S(q)
\rightarrow F(k) + S(l)$.}
\label{fig:1}
\end{figure}
\begin{figure}
\protect
\caption{$P$ as a function of the fermion mass ($m$)
for $E=100$ MeV, $M=1000$ MeV, and $\theta= 20^\circ$ (solid line),
$40^\circ$ (dashed line), $60^\circ$ (dot-dashed line), and $80^\circ$
(dotted line).}
\label{fig:2}
\end{figure}
\begin{figure}
\protect
\caption{$P$ as a function of the scalar mass ($M$) for $E=100$ MeV,
$\theta= 30^\circ$, and $m= 30$ MeV (solid line), $50$ MeV (dashed
line), $70$ MeV (dot-dashed line), and $90$ MeV (dotted line).}
\label{fig:3}
\end{figure}


\begin{references}
\bibitem[*]{byline} Universit\'e Louis Pasteur, Institut Universitaire
de Technologie  3 rue S\b{t} Paul, 67300 Strasbourg, France.

\bibitem{P} A.\ Papapetrou, Proc.\ R.\ Soc.\  London {\bf A209}, 2481
(1951).

\bibitem{W} D.\ R.\ Brill and J.\ A.\ Wheeler, Rev.\ Mod.\ Phys.\
{\bf 29}, 465 (1957).

\bibitem{CP} Y.\ Q.\ Cai and G.\ Papini,  Phys.\ Rev.\ Lett.\
{\bf 66}, 1259 (1991); {\it idem }  {\bf 68}, 3811 (1992).

\bibitem{A} J.\ Anandan. Phys.\ Rev.\ Lett.\ {\bf 68}, 3809 (1992).

\bibitem{DW}  B.\ S.\ DeWitt,  Phys.\ Rev. {\bf 162}, 1239 (1967);
D.\ J.\ Gross and R.\ Jackiw, Phys.\ Rev.\ {\bf 166}, 1287 (1968).

\bibitem{ANS}  V.\ I.\ Ogievetskii, and I.\ V.\ Polubarinov, Soviet
Phys.\ JETP {\bf 21}, 1093 (1965);
S.\ F.\ Novaes, and D.\ Spehler, Phys.\ Rev.\ D{\bf 44}, 3990
(1991); R.\ Aldrovandi, S.\ F.\ Novaes, and D.\ Spehler,  Gen.\
Relativ.\ Gravit. (to appear).

\end{references}
\end{document}